# Complex Networks as Hypergraphs

Ernesto Estrada[1*] and Juan A. Rodríguez-Velázquez[2]

[1]*Complex Systems Research Group*, X-rays Unit, RIAIDT, Edificio CACTUS, University of Santiago de Compostela, 15706 Santiago de Compostela, Spain and
[2]Department of Mathematics, University Carlos III de Madrid, 28911 Leganés (Madrid), Spain.

The representation of complex systems as networks is inappropriate for the study of certain problems. We show several examples of social, biological, ecological and technological systems where the use of complex networks gives very limited information about the structure of the system. We propose to use *hypergraph*s to represent these systems by introducing the concept of the *complex hyper-network*. We define several structural measures for complex hyper-networks. These measures characterize hyper-network structures on the basis of node participation in different *hyper-edges* (groups) and sub-hypergraphs. We also define two clustering coefficients, one characterizing the transitivity in the hyper-network through the proportion of hyper-triangles to paths of length two and the other characterizing the formation of triples of mutually adjacent groups in the hyper-network. All of these characteristics are studied in two different hyper-networks; a scientific collaboration hyper-network and an ecological competence hyper-network.

## 1. INTRODUCTION

The study of complex networks represents an important area of multidisciplinary research involving physics, mathematics, chemistry, biology, social sciences, and information sciences, among others [1–5]. These systems are commonly represented by means of simple or directed graphs that consist of sets of nodes representing the objects under investigation, e.g., people or groups of people, molecular entities, computers, etc., joined together in pairs by links if the corresponding nodes are related by some kind of relationship. These networks include the Internet [6], the World Wide Web [7], social networks [8–11], information networks [12, 13], neural networks [14], food webs [15], reaction and metabolic networks [16], and protein–protein interaction networks [17].

In some cases the use of simple or directed graphs to represent complex networks does not provide a complete description of the real-world systems under investigation. For instance, in a collaboration network represented as a simple graph we only know whether scientists have collaborated or not, but we can not know whether three or more authors linked together in the network were coauthors of the same paper or not. A possible solution to this problem is to represent the collaboration network as a bipartite graph in which a disjoint set of nodes represents papers and another disjoint set represents authors. However, in this case the "homogeneity" in the definition of nodes is lost, because we have certain nodes that represent papers and others that represent authors. In the study of connectivity, clustering and other topological properties, this distinction between two classes of nodes with completely different interpretations may lead to artifacts in the data [18].

A natural way of representing these systems is to use a generalization of graphs known as *hypergraphs* [19, 20]. In a graph a link relates only a pair of nodes, but the edges of the hypergraph — known as hyper-edges — can relate groups of more than two nodes. Thus, we can represent the collaboration network as a hypergraph in which nodes represent authors and hyper-edges represent the groups of authors that have published papers together. Despite the fact that complex weighted networks have been covered in some detail in the physical literature, there are no reports on the use of hypergraphs to represent complex systems. Consequently, we will formally introduce the hypergraph concept as a generalization for representing complex networks and will call them *complex hyper-networks*. The hypergraph concept includes, as particular cases, a wide variety of other mathematical structures that are appropriate for the study of complex networks. For instance, the hypergraph concept is a generalization of the graph concept [19, 20], block design [21], projective plane [22] and affine plane [23]. We will first show some examples of complex systems for which hypergraph representation is necessary. We will subsequently define several topological parameters for the study of complex hyper-networks and will apply them to the study of real-world complex hyper-networks.

## 2. EXAMPLES OF COMPLEX HYPER-NETWORKS

### a) Social Networks

In social networks nodes represent people or groups of people, normally called actors, that are connected by pairs according to some pattern of contact or interactions between them [8]. Such patterns can be of friendship, collaboration, sexual contact, business relationships, etc. There are some cases in which hypergraph representations of the social network are indispensable. These are, for instance, the supra-dyadic transactions in social networks in which it is necessary to consider the coordinated actions of more than two actors, such as a buyer, a seller and a broker. Other examples include the scenarios in which not only the actors taking part in the actions are important, but other factors such as places or times in which the actions taking place are essential to describe such acts. Bonacich et al. [24] have taken such additional characteristics into account in extending the eigenvector centrality for hypergraphs representing supra-dyadic transactions. In such hyper-networks the nodes represent actors related by a common process, which is represented by a hyper-edge, such as a commercial transaction.

**b) Reaction and Metabolic Networks**

A chemical reaction is a process in which a set of chemical compounds known as educts, $E_i$, react in certain *stoichiometric* proportions, $e_i$, to be transformed into a set of other chemical compounds named products, $P_i$, which are produced in certain *stoichiometric* quantities $p_i$:

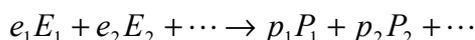

$$e_1 E_1 + e_2 E_2 + \cdots \rightarrow p_1 P_1 + p_2 P_2 + \cdots$$

A chemical reaction can be described as a weighted directed hyper-edge in a directed hypergraph where nodes are the chemicals and hyper-edges are the reactions [25]. The absence of a well-developed theory for the structural analysis of (directed) hypergraphs means that two alternative representations of a chemical reaction are commonly used. The first is the bipartite graph, in which a set of nodes represents educts and products and the other set represents the reaction itself. The other representation consists of the *substrate graph*, which considers educts and products as nodes – two nodes are connected if the corresponding chemical compounds take part in the same reaction.

Metabolic networks can be considered as particular cases of reaction networks that are structurally well-characterized as they can be reconstructed for many organisms up to genome-scale. Metabolic pathways are represented in the form of graphs in which nodes represent molecular entities and edges represent reactions or processes relating the molecules involved in the reaction. However, since a reaction may have more than one substrate, and more than one product, the pathway is better represented by a hyper-network in which hyper-edges represent reactions and nodes represent molecular entities [26]. As reaction graphs, metabolic networks are normally represented as substrate or bipartite graphs. Some problems arise when these representations are used for the analysis of potential failure modes in the metabolic network [27].

**c) Protein Complex Networks**

The systematic characterization of multi-protein complexes in the whole proteome of an organism requires the data to be organized in the form of protein membership lists of the protein complexes. The most common forms of this organization are the protein-protein interaction networks and the complex intersection graphs. In the first representation the nodes of the network represent proteins and an edge links two proteins that interact with each other. This representation, however, does not take into account the multi-protein complexes. In the complex intersection graph the nodes represent complexes, and a link exists between two complexes if they have one or more proteins in common. Clearly, this second representation does not provide information about proteins. A natural way of accounting for the information about both proteins and common protein membership in the complexes, such as common regulation, localization, turnover, or architecture, is to use a hypergraph representation. In the protein complex hyper-networks each protein is represented by a node and each complex by a hyper-edge [28]. These kinds of hyper-networks can be visualized as bipartite graphs.

**d) Food Webs**

Trophic relations in ecological systems are normally represented through the use of food webs, which are oriented graphs (digraphs) whose nodes represent species and links represent trophic relations between species [29]. Another way of representing food webs is by means of competition graphs $C(G)$, which have the same set of nodes as the food web but in which two nodes are connected if, and only if, the corresponding species compete for the same prey in the food web [30]. In the competition graph we can only know if two linked species have some common prey, but we can not know the composition of the whole group of species that compete for common prey. In order to solve this problem a competition hypergraph has been proposed in which nodes represent species in the food web and hyper-edges represent groups of species that compete for common prey [31]. It has been shown that in many cases competition hyper-networks yield a more detailed description of the predation relations among the species in the food web than competition graphs. A food web and its competition network and hyper-network are illustrated in Fig. 1.

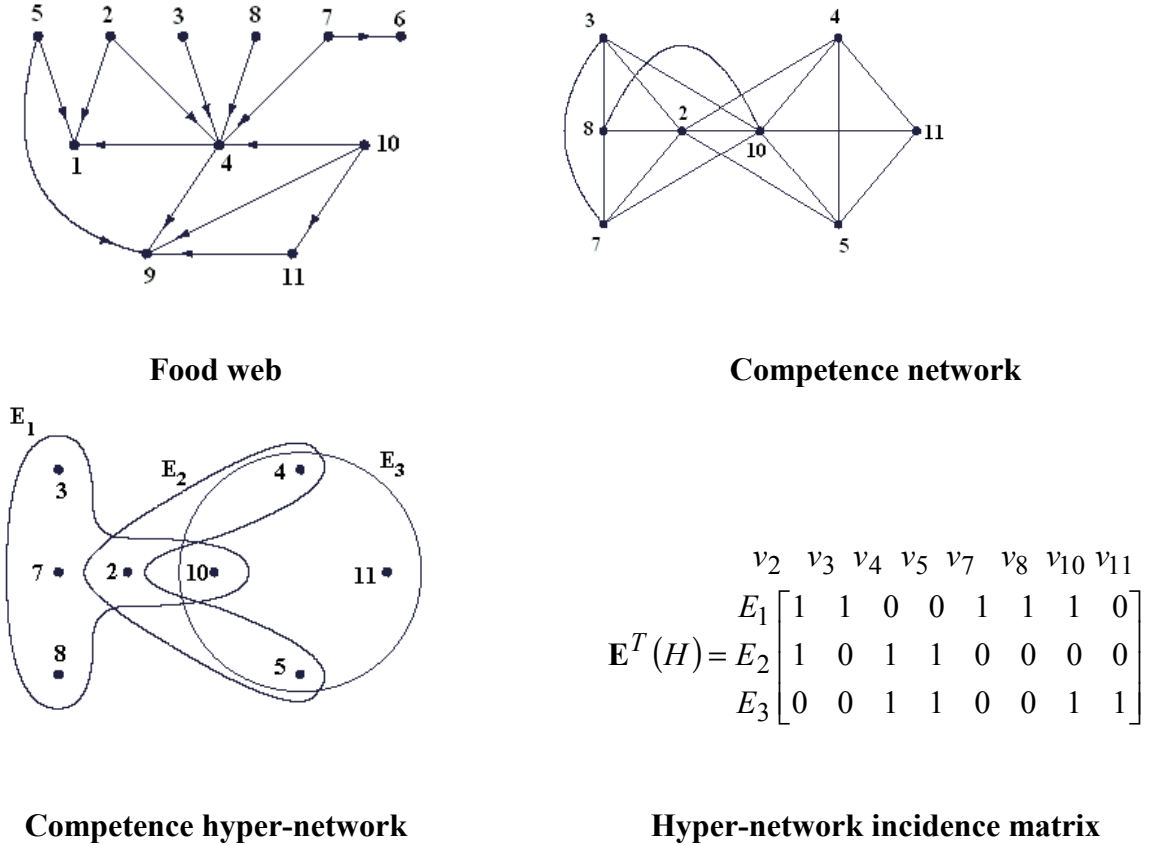

**Food web**                                    **Competence network**

**Competence hyper-network**            **Hyper-network incidence matrix**

$$\mathbf{E}^T(H) = \begin{array}{c} \\ E_1 \\ E_2 \\ E_3 \end{array} \begin{array}{cccccccc} v_2 & v_3 & v_4 & v_5 & v_7 & v_8 & v_{10} & v_{11} \\ \left[\begin{array}{cccccccc} 1 & 1 & 0 & 0 & 1 & 1 & 1 & 0 \\ 1 & 0 & 1 & 1 & 0 & 0 & 0 & 0 \\ 0 & 0 & 1 & 1 & 0 & 0 & 1 & 1 \end{array}\right] \end{array}$$

FIG 1. Food web for a Malaysian rain forest, its competition network and hyper-network and the transposed incidence matrix representing the hyper-network.

**d)   Other complex systems**

There are other fields in which hypergraphs have been used to study complex systems and there are many others in which the use of hyper-networks may provide an interesting alternative. Harn *et al*. [32] used hyper-networks to study the software evolution process in which an *evolutionary hypergraph* was defined as a labeled, directed and acyclic hypergraph. A different application of hypergraphs was developed by Krackhardt and Kilduff, who studied the so-called Simmelian tied dyads, which are dyads embedded in three-person cliques, in three entrepreneurial firms [33]. Hypergraphs have also appeared as a natural consequence of an L-percolation process in complex networks, as studied by da Fontoura Costa [34], as well as in the detection of hidden groups in communication networks [35]. All of these applications clearly indicate the importance of hypergraphs for representing and studying complex systems.

## 3. HYPERGRAPH DEFINITIONS

A *hypergraph* is represented by a pair $H = (V, E)$, where $V = \{v_1, v_2, \cdots, v_N\}$ is the set of *nodes* and $E = \{E_1, E_2, \cdots, E_m\}$ is the set of *hyper-edges*, such that $E_i \neq \varnothing$ and $\bigcup_i E_i = V$. A hyper-edge $E_i$ is a sub-set of $V$ [19, 20]. Two nodes are adjacent in $H = (V, E)$ if there is a hyper-edge $E_i$ that contains both of these nodes. A simple hypergraph is a hypergraph $H$ such that $E_i \subseteq E_j \Rightarrow i = j$. A simple graph is a simple hypergraph, each edge of which has cardinality 2 [19, 20]. We will refer to *complex hyper-networks* as hypergraphs that represent a complex system, such as those previously described in this work.

A hypergraph $H$ can be represented by an incidence matrix $E(H) = E = \lfloor e_{ij} \rfloor$ such that $e_{ij} \in \{0,1\}$, in which each of the $N$ rows is associated with a vertex and each of the $m$ columns is associated with a hyper-edge, where $e_{ij} = 1$ if $v_i \in E_j$ and $e_{ij} = 0$ if $v_i \notin E_j$. Thus, any Boolean matrix may be considered as the incidence matrix of a hypergraph. The degree of the node $v$ is the sum of the corresponding row of the incidence matrix of $H$. This is equal to the number of hyper-edges that the node belongs to [19].

The adjacency matrix, $\mathbf{A}(H)$, of the hypergraph $H = (V, E)$ is a square symmetric matrix whose entries $a_{ij}$ are the number of hyper-edges that contain both nodes $v_i$ and $v_j$: the diagonal entries of $\mathbf{A}(H)$ are zero. This can be obtained from the incidence matrix of $H$ as follows:

$$\mathbf{A}(H) = \mathbf{E}\mathbf{E}^T - \mathbf{D}$$

where $\mathbf{E}^T$ is the transpose of the incidence matrix and $\mathbf{D}$ is the diagonal matrix whose diagonal entries are the degrees of the vertices. More formally, $\mathbf{A}(H)$ is a $|V| \times |V|$ matrix with diagonal entries $A_{ii} = 0$, for $v_i \in V$, and off-diagonal entries

$$A_{ij} = \left| \left\{ E_k \in E : \{v_i, v_j\} \subset E_k \right\} \right|, \text{ for } v_i, v_j \in V, \ i \neq j.$$

Since $\mathbf{A}(H)$ is symmetric, and its entries are non-negative integers, it may be viewed as the adjacency matrix of a multigraph $G'$, i.e., a graph having multiple links between nodes, called the *associated graph* of $H = (V, E)$ [36]. Consequently, all those topological parameters of a hyper-network that are derived from its adjacency matrix are numerically "identical" to those derived from the corresponding associated multigraph. However, it should be noted that multigraphs, or more generally weighted networks, and hyper-networks are two different ways of representing complex systems and they provide different information about the topology of such systems. Even the information given by the associated multigraph of a hypergraph is not as complete as that given by the proper hyper-network. For instance, on using the associated multigraph we can not say whether a group of nodes are related by the same relation (multiedge) or not, i.e., the associated multigraph is not a supra-dyadic representation of the complex system.

A *walk* of length $l$ in $H = (V, E)$ is defined to be a sequence of (not necessarily different) vertices $(v_1, v_2, \cdots, v_l, v_{l+1})$ such that for each $i = 1, 2, \ldots, l$ there is a hyper-edge containing $v_i$ and $v_{i+1}$. The walk is *closed* (CW) if $v_{l+1} = v_1$. A *path* is a walk in which all vertices and hyper-edges are distinct. A *cycle* is a CW in which all vertices and hyper-edges are distinct.

For a set $J \subset \{1, 2, \ldots, m\}$ we call the family $H_{(V, J)} = (E_j : j \in J)$ the *partial hypergraph of $H = (V, E)$ generated by the set $J$*. For a set $A \subset V$ we call the family $H_{(A, E)} = (E_j \cap A : 1 \leq j \leq m, E_j \cap A \neq \phi)$ the *sub-hypergraph of $H = (V, E)$ induced by the set* of vertices $A$. Partial hypergraphs and sub-hypergraph induced by sets of vertices are both particular cases of a *sub-hypergraph* of $H = (V, E)$. In general, for a set $J \subset \{1, 2, \ldots, m\}$ and a set $A \subset V$ we call the family $H_{(A, J)} = (E_j \cap A : j \in J, E_j \cap A \neq \phi)$ the *sub-hypergraph* of $H = (V, E)$ induced by the sets $A$ and $J$.

The dual of a hypergraph $H = (V, E)$ is a hypergraph $H^* = (V^* = E, E^*)$, whose vertices correspond to the edges of $H$ with edges $E_i^* = \{E_j : v_i \in E_j \text{ in } H\}$. Clearly, the incidence matrix of $H^*$ is the transpose of the incidence matrix of $H$ and so we have $\left(H^*\right)^* = H$. The reader is referred to the literature for more details on hypergraphs [19, 20].

## 4. SUB-HYPERGRAPH CENTRALITY

Let $H$ be a simple hypergraph of order $N$. Since the adjacency matrix, $A$, of $H$ is a symmetric matrix with real entries, there exists an orthogonal matrix $U = (u_{ij})$ such that $A = U D U^T$, where $D = diag(\lambda_1, \lambda_2, \ldots, \lambda_N)$ whose diagonal entries are the eigenvalues of $A$, and the columns of $U$ are the corresponding eigenvectors that form an orthogonal basis of the Euclidean space $\Re^N$. It must be emphasized that, if the hypergraph $H$ is connected, then the symmetric and non-negative matrix $A$ is irreducible. As a consequence, the main eigenvalue $A$ has a positive eigenvector of multiplicity one. This fact facilitates the extension, to the case of hypergraphs, of the use of the main eigenvector as a measure of centrality. The following result that was obtained in [37] will be useful in extending the definition of subgraph centrality [38] to hypergraphs.

**Theorem 1:** *Let $v_i$ and $v_j$ be vertices of a hypergraph $H$. Let $A$ be the adjacency matrix of $H$. Then, the number of walks of length $k$ in $H$, from $v_i$ to $v_j$, is the entry in position $(i, j)$ of the matrix $A^k$.*

From the above theorem we can see that walks of length $k$ in $H$, from $v_i$ to $v_j$, are $\mu_k(ij) = \left(A^k\right)_{ij} = \sum_{s=1}^{N} u_{is} u_{js} \lambda_s^k$.

Hence, the number $W_k$ of walks of length $k$ in $H$ is given by

$$W_k = \sum_{i,j} \mu_k(ij) = \sum_{s=1}^{N} \left( \sum_{i=1}^{N} u_{is} \right)^2 \lambda_s^k .$$

Moreover, the number of closed walks of length $k$ starting and ending on vertex $v_i$ in $H$ is given by the local spectral moments $\mu_k(i)$, which are simply defined as the $i$ th diagonal entry of the $k$ th power of the adjacency matrix, $\mathbf{A}$:

$$\mu_k(i) = \left( \mathbf{A}^k \right)_{ii} = \sum_{s=1}^{N} (u_{is})^2 \lambda_s^k , \tag{1}$$

and the number $CW_k$ of closed-walks of length $k$ in $H$ is given by

$$CW_k = \sum_{i} \mu_k(i) = \sum_{s=1}^{N} \lambda_s^k , \text{ i.e., the trace of } A^k .$$

We define the *sub-hypergraph centrality* of the vertex $v$ as the "sum" of closed walks of different lengths in the network starting and ending at vertex $v$. As this sum includes both trivial and non-trivial closed walks, we must consider all sub-hypergraphs, i.e., acyclic and cyclic. The contribution of these closed walks decreases as the length of the walks increases. In other words, shorter closed walks have more influence on the centrality of the vertex than longer closed walks. This rule is based on the observation that motifs in real-world networks are small sub-hypergraphs. The extreme case is that of closed walks with a length of only two, giving a weight of zero to longer walks. This case corresponds to a vertex "degree centrality" $d_V(v_i)$ defined as the number of vertices adjacent to $v_i$, that is, $d_V(v_i) = \left| \left\{ v_j : v_j \sim v_i \right\} \right|$. On the other hand, the use of the sum of closed walks to define sub-hypergraph centrality presupposes a mathematical problem as the series $\sum_{k=0}^{\infty} \mu_k(i) = \infty$ diverges. Consequently, we avoid this problem by scaling the contribution of closed walks to the centrality of the vertex by dividing them by the factorial of the order of the spectral moment. The *sub-hypergraph centrality* of vertex $v_i$ in the network is then given by:

$$C_{SH}(i) = \sum_{k=0}^{\infty} \frac{\mu_k(i)}{k!} . \tag{2}$$

Let $\lambda$ be the main eigenvalue of $\mathbf{A}$. For any non-negative integer $k$ and any $i \in \{1, ..., n\}$, $\mu_k(i) \leq \lambda^k$, series (2) – whose terms are non-negative – converges.

$$\sum_{k=0}^{\infty} \frac{\mu_k(i)}{k!} \leq \sum_{k=0}^{\infty} \frac{\lambda^k}{k!} = e^{\lambda} \tag{3}$$

Thus, the sub-hypergraph centrality of any vertex $v_i$ is bounded above by $C_{SH}(i) \leq e^{\lambda}$. The following result shows that the sub-hypergraph centrality can be obtained mathematically from the spectrum of the adjacency matrix of the network.

**Theorem 2:** *Let* $H = (V, E)$ *be a simple hypergraph of order* $N$. *If* $v_i \in V$, *then the sub-hypergraph centrality* $C_{SH}(i)$ *may be expressed as follows*:

$$C_{SH}(i) = \sum_{j=1}^{N} (u_{ij})^2 e^{\lambda_j} \qquad (4)$$

*Proof*: Using expressions (1) and (2), we obtain

$$C_{SH}(i) = \sum_{k=0}^{\infty} \left( \sum_{j=1}^{N} \frac{\lambda_j^k (u_{ij})^2}{k!} \right). \qquad (5)$$

By reordering the terms of series (7), we obtain the absolutely convergent series:

$$\sum_{j=1}^{N} \left( (u_{ij})^2 \sum_{k=0}^{\infty} \frac{\lambda_j^k}{k!} \right) = \sum_{j=1}^{N} \left( (u_{ij})^2 e^{\lambda_j} \right), \qquad (6)$$

which, clearly, also converges to $C_{SH}(i)$, proving the main result.

A global characterization of the network $H$ can be carried out by obtaining the mean of the average sub-hypergraph centrality: $\langle C_{SH} \rangle = \frac{1}{N} \sum_{i=1}^{N} C_{SH}(i)$. It has been recommended that the use of centralization instead of centrality is more appropriate for these sorts of global measures [8]. An analytical expression for $\langle C_{SH} \rangle$ can be obtained using a procedure analogous to that described to prove the previous theorem, showing that $\langle C_{SH} \rangle$ depends only on the eigenvalues and size of the adjacency matrix of the network:

$$\langle C_{SH} \rangle = \frac{1}{N} \sum_{i=1}^{N} C_{SH}(i) = \frac{1}{N} \sum_{i=1}^{N} e^{\lambda_i} \qquad (7)$$

## 5. CLUSTERING COEFFICIENT

One of the most important topological parameters used in the study of complex networks is the clustering coefficient. The clustering coefficient measures the degree of cliquishness that a network has. Watts and Strogatz defined a local clustering coefficient that describes "what proportion of acquaintances of a vertex know each other" [14]. In this respect, the global clustering of a network is obtained as the average of the local clustering coefficients for all nodes in the network. It has been stated that this kind of "average of an average" [39] is often not very informative and that a better alternative is to use the following definition of clustering coefficient for a network, which is also known in the sociology literature as the transitivity coefficient [8]:

$$C_2(G) = \frac{3 \times number\ of\ triangles}{number\ of\ pairs\ of\ adjacent\ edges} = \frac{6 \times number\ of\ triangles}{number\ of\ paths\ of\ length\ two} \qquad (8)$$

The factor of three in the numerator compensates for the fact that each triangle contributes three connected triples and ensures that $C_2(G) = 1$ for the complete graph $K_N$. In the case of multigraphs, i.e., graphs with multiple edges, this proportion is not maintained and the multigraph is represented as a simple graph to calculate the clustering coefficient. A similar situation is presented for hypergraphs. Thus, in those cases in which the hypergraph has multi-hyper-edges we will consider them as simple hyper-edges. More formally, we define an associated simple graph $G_H$ to the hyper-network $H$, in which two nodes of $G_H$ are adjacent if, and only if, they are adjacent in $H$. This is equivalent to removing the multiple links between vertices in $H$. In order to account for the cliquishness of a hypergraph we have to modify (8) to the following expression:

$$C_2(H) = \frac{6 \times number\ of\ hyper\text{-}triangles}{number\ of\ 2 - paths}, \qquad (9)$$

where a hyper-triangle is defined as a sequence of three different vertices and three different hyper-edges of the form: $v_i, E_p, v_j, E_q, v_k, E_r, v_i$, in which the three nodes are mutually adjacent and a 2-path is path of length 2, i.e., a sequence of the type $v_i, E_p, v_j, E_q, v_k$ (we recall that in a path all vertices and hyper-edges are distinct).

We will use the associate graph $G_H$ to compute the clustering coefficient of $H$. We can use the number of closed walks of length three in $G_H$ to count the number of hyper-triangles in the hyper-network. However, we have to exclude those CWs of length three that are not hyper-triangles. For instance, in Fig. 1 it is shown that $v_2, E_2, v_4, E_3, v_{10}, E_1, v_2$ is an example of hyper-triangle but neither $v_2, E_2, v_4, E_2, v_5, E_2, v_2$ nor $v_2, E_2, v_4, E_3, v_5, E_2, v_2$ are hyper-triangles despite the fact that they are CWs of length three. The CWs of length three that are not hyper-triangles come from the nodes that are on the same hyper-edge. For the sake of simplicity in the terminology we will call them "false" hyper-triangles.

The number of CW of length three containing only one hyper-edge $E_i$ is given by $t_i = \begin{pmatrix} |E_i| \\ 3 \end{pmatrix}$. These are the number of CWs of length three formed inside a single hyper-edge and, consequently, are not hyper-triangles, e.g., $v_2, E_2, v_4, E_2, v_5, E_2, v_2$ in Fig. 1. In general, to calculate the number of false hyper-triangles we need to use the inclusion-exclusion principle. The cardinal of the intersection of hyper-edges $E_{i_1}, E_{i_2}, ..., E_{i_k}$, $\alpha_{i_1 i_2 ... i_k} = \left| \bigcap_{r=1}^{k} E_{i_r} \right|$ is denoted by $\alpha_{i_1 i_2 ... i_k}$. The number of false hyper-triangles is then $t = \sum_{j=1}^{m} (-1)^{j+1} a_j$, where $a_k = \sum_{i_1, i_2 ... i_k} \begin{pmatrix} \alpha_{i_1 i_2 ... i_k} \\ 3 \end{pmatrix}$ (we only consider the terms in which the combinatorial expression makes sense).

On the other hand, we have to count the number of 2-paths in the hyper-network, which is the denominator of the expression of the clustering coefficient (11). In this respect, we need to identify the number of walks of length two between nodes in the same hyper-edge $E_i$ because they do not constitute a path of length two, i.e., they are "false" 2-paths. The number of false 2-paths is $p = 3t$, where $t$ is the number of false hyper-triangles. The clustering coefficient of the hyper-network is now given by:

$$C_2(H) = \frac{CW_3(G_H) - 6t}{W_2(G_H) - CW_2(G_H) - 6t} = \frac{\sum_i \mu_3(i) - 6t}{W_2(G_H) - \sum_i \mu_2(i) - 6t} \qquad (10)$$

which, by substitution, gives the expression of the clustering coefficient for a hyper-network as presented in the following theorem.

**Theorem 3:** Let $t$ *be the number of false hyper-triangles of a hyper-network* $H$. *Let* $\lambda_1, \lambda_2, ..., \lambda_N$ *be the eigenvalues of* $G_H$ *and let* $U = (u_{ij})$ *denote an orthogonal matrix whose columns are the corresponding eigenvectors, which form an orthogonal basis of the Euclidean space* $\Re^N$. *The clustering coefficient of* $H$ *is given by:*

$$C_2(H) = \frac{\sum_{s=1}^{N} \lambda_s^3 - 6t}{\sum_{s=1}^{N} \left( \left( \sum_{j=1}^{N} u_{is} \right)^2 - 1 \right) \lambda_s^2 - 6t} \qquad (11)$$

The clustering coefficient for the competition network in Fig. 1 is $C_2(H) = 0.25$, which indicates that 1/4 of the 18 triples of nodes participating in at least two different competence groups do participate in three different groups. They are $v_2, v_4, v_{10}$ and $v_2, v_5, v_{10}$, which form the hyper-triangles $v_2, E_2, v_4, E_3, v_{10}, E_1, v_2$ and $v_2, E_2, v_5, E_3, v_{10}, E_1, v_2$, respectively.

It is feasible that we could be interested in knowing the proportion of triples of groups that are mutually adjacent in the hyper-network forming triangles with respect to the number of triples of groups that only form two pairs of adjacent groups. For instance, if we consider three different groups $E_1, E_2, E_3$ in a hyper-network it is possible to form three pairs of adjacent groups: $E_1, E_2$, $E_1, E_3$ and $E_2, E_3$ and only one triangle of mutually adjacent groups. If we now define the clustering coefficient of the dual of the hyper-network we will obtain the proportion of triples of groups forming triangles with respect to the number of triples forming adjacent groups:

$$C_2(H^*) = \frac{3 \times number\ of\ triangles}{number\ of\ pairs\ of\ adjacent\ edges} \qquad (12)$$

which in terms of the graph spectrum is given by the following expression:

$$C_2(H^*) = \frac{\sum_{s=1}^{m} \mu_s^3}{\sum_{s=1}^{m} \left( \left( \sum_{j=1}^{m} b_{is} \right)^2 - 1 \right) \mu_s^2} \qquad (13)$$

where $\mu_1, \mu_2, ..., \mu_m$ are the eigenvalues of the simple graph $G_{H^*}$ and $B = (b_{ij})$ denotes an orthogonal matrix whose columns are the corresponding eigenvectors that form an orthogonal basis of the Euclidean space $\Re^m$.

In the competition hyper-network represented in Fig. 1 the values of $C_2(H^*) = 1$, which indicates that the three competence groups in the network are mutually adjacent. In this case the trophic species **2** predates together with competitors in the competition groups $E_1, E_2$; species **4** and **5** participate in the competition groups $E_2, E_3$, and species **10** competes with predators in groups $E_1, E_3$.

## 6. ANALYSIS OF REAL-WORLD HYPER-NETWORKS

We will consider here two complex hyper-networks representing a collaboration network and a competence graph in an ecological system. The collaboration network was extracted from the bibliography of the book "Product Graphs" by Imrich and Klavžar [40]. The original network is a bipartite author-by-paper network where link (*i,j*) represents a situation where author *i* is the (co)author of the paper *j* [41]. We transformed this information into a complex hyper-network in which nodes are authors and hyper-edges are papers in such a way that all coauthors of a paper are linked by the same hyper-edge. The second network consists of the trophic relation between species in the marine ecosystem of Benguela, which is off the southwest coast of South Africa [42]. We represent this network as a competence graph, as explained before, and in this way we obtain the Benguela competence hyper-network in which nodes are species and hyperlinks join together all species competing for the same prey. For the sake of comparison we also consider the author-by-author complex networks, representing those pairs of authors that have published a joint paper, as well as the competence graph of the Benguela food web.

We begin by analyzing the characteristics of the degree of centrality in the hyper-networks studied and compare them with the corresponding network representations of the same complex systems. One of the distinctive characteristics of centrality measures is that they allow the nodes of the network to be ranked in order to determine the most central ones in such a system. The relative degree of centrality for the most central authors in the "Graph Product" collaboration network are plotted in Fig. 2A according to three different representations of the system: network, weighted network and hyper-network. As can be seen, the ranking obtained for the authors is completely different for the three representation methods. It was found that Hell is the most central author in the simple network, whereas Klavžar and Imrich are more prominent in the weighted network and in the hyper-network, respectively. There are some other authors that appear highly ranked in one of the representations but do not appear among the top ten authors in the others. For instance, Zhu is ranked as the fifth author according to the hyper-network representation but does not appear among the top ten authors in the network and weighted network. Imrich and Klavžar are the most central authors in the complex hyper-network. This means that they appear in the largest number of hyper-edges, which represent the different collaboration groups in the hyper-network. We give the term *collaboration group* to the set of authors that collaborate together in a paper. Imrich and Klavžar participate in 26 and 22 collaboration groups, respectively, while Hell participates in only 8 collaboration groups – despite the fact that he has 12 collaborators. It is clear that you can have N collaborators but participate in only one collaboration group if all your collaborators participate in the same paper. The importance of participating in different collaboration groups is evident from the following perspective. If your N collaborators are in only one group and all of them can share the same series of ideas, i.e., they form a "school". However, if you are taking part in two or more collaboration groups you will be in touch with more than one school of thought and share ideas from different perspectives. This makes nodes participating in larger numbers of collaboration groups (hyper-edges) the most central nodes in the hyper-network.

A similar situation is represented in Fig. 2B, where the relative degrees based on simple network and hyper-network are plotted for all species in the competence network and hyper-network of the Benguela ecosystem. In the competence network there are 10 species ranked as the most central ones, all of which have a degree equal to 22. This means that each of these species competes for 22 types of prey. For instance, anchovy, horse and chub mackerel all compete for the same number of prey as sharks and birds in that they all have a centrality degree equal to 22 in the competence network. The possibility exists that most of the prey for which a group of species compete are the same. We assign the term *competence group* to this group of species that compete for the same prey. The competence network does not allow us to know in how many competence groups a particular species is participating. However, this information can be obtained in a straightforward way from the competence hyper-network. In this case the node degree corresponds to the number of competence groups in which a species is participating. In the hyper-network representation of the Benguela ecosystem, sharks are the most central species followed by birds and seals. Sharks participate in 18 different competition groups while birds participate in 16 and seals in 15. In contrast, anchovy, horse and chub mackerel participate in only 4 competence groups, which makes sharks, birds and seals the most central species in the competence hyper-network.

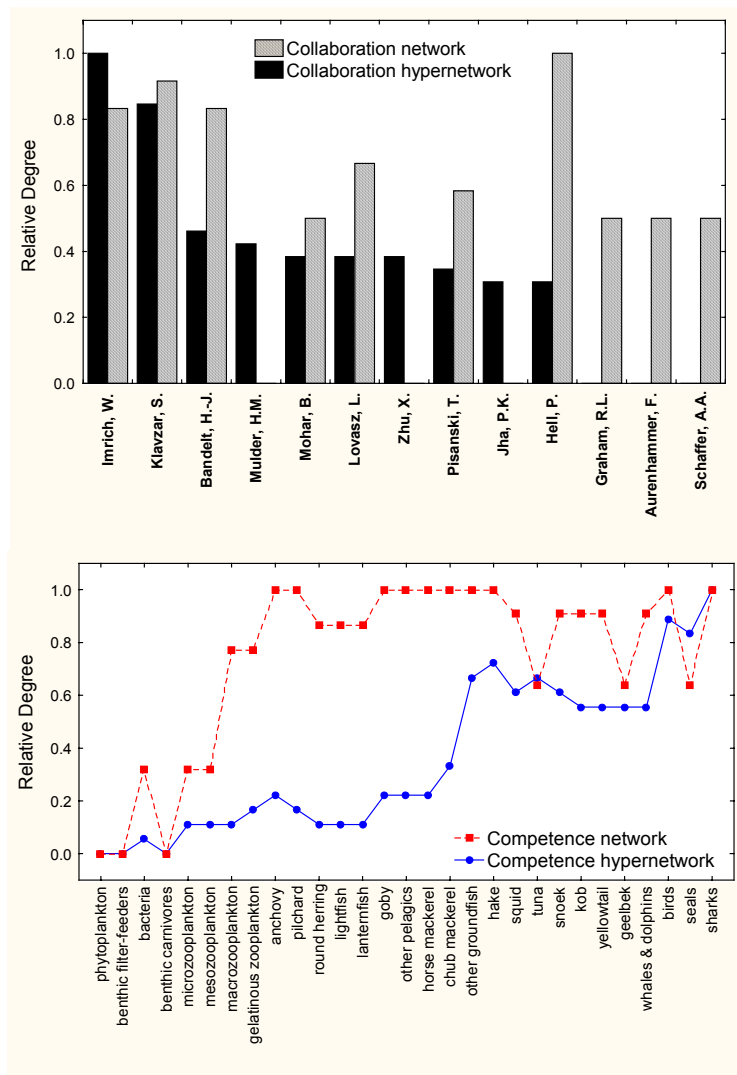

FIG 2. Ranking of nodes according to the relative degree centrality. Authors in the collaboration network and hyper-network of "Graph Product" (top) and species in the competition network and hyper-network for a Malaysian rain forest food web (bottom).

We also studied the sub-hypergraph centrality of complex hyper-networks by calculating $C_{SH}(i)$ and $\langle C_{SH} \rangle$ for the two hyper-networks studied here as well as their equivalent values for the corresponding simple networks. In the "Product Graphs" collaboration hyper-network there are significant differences in the ranking of nodes compared to that observed in the collaboration network. $C_{SH}(i)$ in the hyper-network ranks Klavžar as the most central author, followed by Imrich, Mohar and Gutman (see Fig. 3A). Hell, who is the most central author in the collaboration network, is not among the top ten authors in the hyper-network. In general, 50% of the authors ranked in the top ten most central nodes in the hyper-network do not appear in the network and vice-versa. On the other hand, there are significant differences between the ranking introduced by $C_{SH}(i)$ in the hyper-network and that obtained by node degrees – as can be seen by comparing Figs. 2A and 3A.

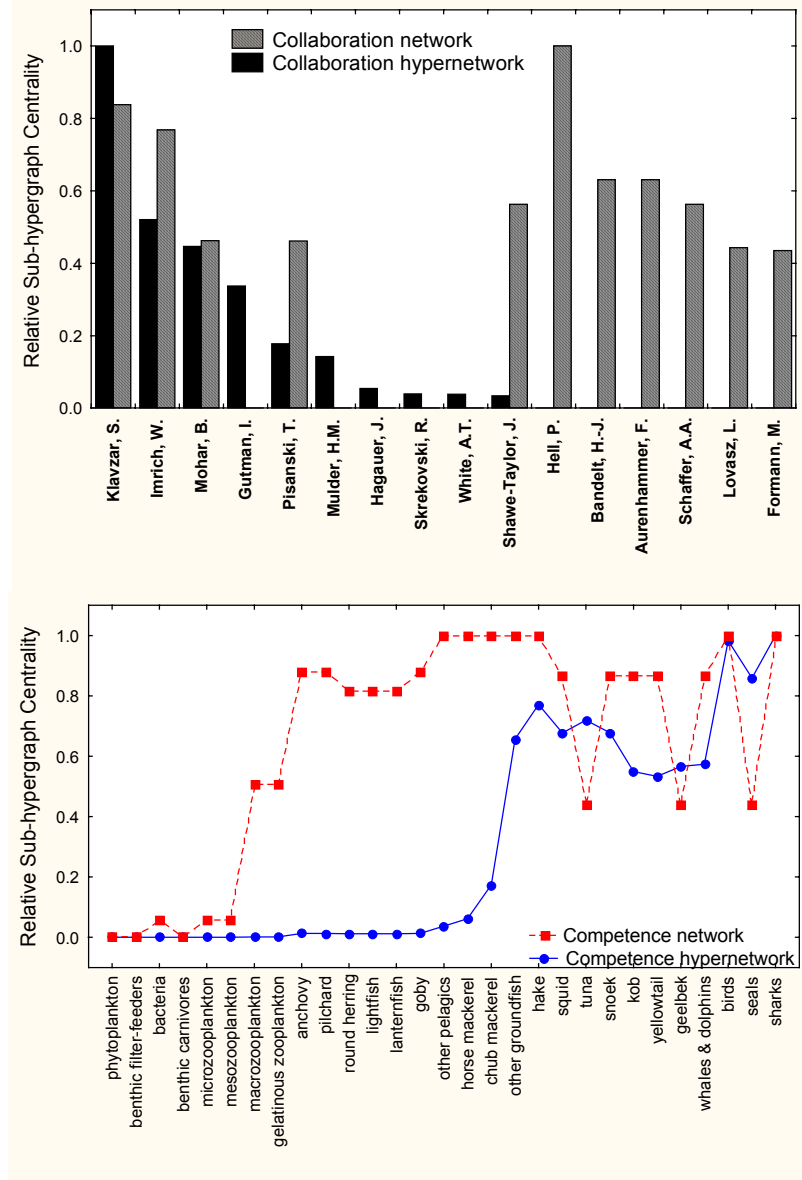

FIG 3. Ranking of nodes according to the relative subgraph centrality. Authors in the collaboration network and hyper-network of "Graph Product" (top) and species in the competition network and hyper-network for a Malaysian rain forest food web (bottom).

In the Benguela ecosystem the competence hyper-network clearly identifies sharks, birds and seals as the most central species according to the sub-hypergraph centrality. In the competence network there are 7 species that are ranked as the most central ones. They include sharks and birds – but not seals – along with horse and chub mackerel, hake, other pelagics and other groundfish.

The clustering coefficient of the Benguela competence hyper-network is $C_2(H) = 0.067$ because there are 14 hyper-triangles and 625 paths of length two. Only two hyper-triangles exist in the "Graph Product" collaboration network, which has a clustering of $C_2(H) = 0.016$ (there are 377 paths of length 2). These values indicate a low transitivity in the hyper-network as the number of hyper-triangles formed is low compared to the number of paths of length 2. Each hyper-triangle in a hyper-network represents a triple of nodes that join together three different groups (hyper-edges). For instance, the hyper-triangle $v_2, E_2, v_4, E_3, v_{10}, E_1, v_2$ is formed by three trophic species ($v_2, v_4, v_{10}$) that join together three different competence groups ($E_1, E_2, E_3$). Thus, the clustering coefficient of the hyper-network $C_2(H)$ measures the proportion of triples of nodes that join three different groups with respect to the number of triples that only join two different groups. With the aim of extracting more conclusive results concerning the role played by transitivity in complex hyper-networks, we propose the further study of random hyper-networks in order to show whether real-world hyper-networks show, for instance, "small-world" characteristics as observed for complex networks. A value of $C_2(H*) = 0.963$ is obtained for the competition graph of the Benguela ecosystem, which is close to 1. This result indicates a high level of interrelationship between the different competition groups in these ecosystems. An analysis of this factor for a greater dataset of food webs is necessary to obtain definitive conclusions about the role of this interrelation in ecological systems. The collaboration network on "Product Graphs" also shows a high value of the clustering coefficient between collaborating groups, $C_2(H*) = 0.758$. This result is not unexpected given the limited scope of the collaboration topic, which makes the different groups working in the field collaborate to a large extent.

## 7. CONCLUSIONS

We have introduced here some basic principles for the use of more general representations of complex systems based on hypergraphs. We have coined the term *complex hyper-networks* to designate such systems in which nodes are grouped together in multi-dyadic relationships represented by hyper-edges. The use of complex hyper-networks appears to be a necessity for exploring several social, technological, biological and ecological systems beyond the traditional dyadic representation of node-node relationships. We have introduced several valuable measures for studying complex hyper-networks, such as node and sub-hypergraph centralities as well as clustering coefficients for both hyper-networks and their duals. The application of these measures to the study of two real-world complex systems – representing a collaboration hyper-network and a competition hyper-network in an ecological system – has shown some of the main differences between complex networks and hyper-networks. However, a large number of open questions remain concerning the representation of complex networks as hypergraphs. For instance, are complex hyper-networks "small-worlds"? Do complex hyper-networks show "scale-free" characteristics? What is the large-scale structural organization of complex hyper-networks? How robust are complex hyper-networks to random failures and attacks? We hope that the current work encourages the investigation of these and many other questions and that the main tools developed here can help in this area.